\documentclass[twocolumn]{revtex4-1}

\def\be{\begin{equation}}
\def\ee{\end{equation}}
\def\la{\label}
\def\bea{\begin{eqnarray}}
\def\eea{\end{eqnarray}}

\def\ci{\cite}
\def\la{\label}
\def\bib{\bibitem}

\def\lm{\lambda}

\def\le{\left}
\def\ri{\right}

\def\p{\phi}
\def\pd{\phi_{24}}

\def\wp{w_\phi}

\def\Ompo{\Omega_{\phi o}}

\def\s8{\sigma_8}

\def\fr{\frac}
\def\pp{\partial}
\def\pu{\pp_\mu}
\def\pU{\pp^\mu}

\def\vp{\varphi}

\def\Op{\Omega_\phi}
\def\Opo{\Omega_{\phi o}}

\def\Odmo{\Omega_{dm o}}

\def\rdmo{\rho_{dm o}}
\def\rdm{\rho_{dm}}

\def\rp{\rho_\phi}

\def\rpo{\rho_{\phi o}}

\def\wp{w_\phi}

\def\Ompo{\Omega_{\phi o}}

\def\G{\Gamma}

\def\Gpff{\Gamma_{\p\psi\psi}}
\def\Gvff{\Gamma_{\vp\psi\psi}}

\def\lin{L_{int}}
\def\vin{V_{int}}

\def\vip{V_{int,\vp}}

\def\mmpo{m^2_{\phi o}}

\def\lin{L_{int}}
\def\mp{m_{\phi}}
\def\mpo{m_{o}}
\def\mpl{m_{pl}}

\usepackage{makeidx}
\usepackage{ctable}
\usepackage{multirow}
\usepackage{graphicx}
\usepackage{grffile}
\usepackage{epstopdf}
\usepackage{subfig}
\usepackage{subfloat}
\usepackage{colortbl}
\usepackage{appendix}
\usepackage{rotating}
\usepackage{array}
\usepackage{caption}

\begin{document}

\title{Dark Matter from the Inflaton Field}
\author{A. de la Macorra }
\affiliation{Instituto de F\'{\i}sica, Universidad Nacional
Autonoma de Mexico, 04510, M\'exico D.F., M\'exico\\
Part of the Collaboration Instituto Avanzado de Cosmologia}
%\thanks{Part of the Instituto Avanzado de Cosmologia}
% \email{macorra@fisica.unam.mx}

\begin{abstract}

We  present a model where inflation and Dark Matter takes place via a single
scalar field $\phi$.  Without
 introducing  any new parameters  we are able unify
inflation and Dark Matter using a scalar field $\p$ that
accounts for inflation at an early epoch  while it gives a
Dark Matter WIMP particle at low energies.
After inflation our universe must be reheated and we must have  a long period of radiation
dominated  before the epoch of Dark Matter. Typically the inflaton decays while it oscillates around
the minimum of its  potential.
If the inflaton decay is not complete or sufficient
then the remaining energy density of the inflaton after reheating
must be fine tuned to give the correct amount of  Dark Matter.
An essential feature here, is that Dark Matter-Inflaton particle
is produced at low energies  without fine tuning or new
parameters. This process uses the same coupling $g$ as for the inflaton decay.
Once the field $\p$ becomes non-relativistic it  will decouple as any WIMP particle, since
$n_\p$ is exponentially suppressed. The correct amount of Dark Matter
determines the cross section  and we have  a constraint between the coupling $g$ and the mass
$m_o$ of $\p$.
The unification scheme  we present  here has four free parameters,
two for the scalar potential $V(\p)$ given by  the inflation parameter $\lm$ of the quartic term
and the mass  $m_o$. The other two parameters
are the  coupling $g$ between the inflaton $\p$ and a scalar filed $\vp$ and the coupling
$h$ between  $\vp$ with standard model particles $\psi$ or $\chi$.
These four  parameters are already present in models of inflation and reheating
process, without considering Dark Matter. Therefore,  our unification scheme does not increase the number
of parameters and it accomplishes the desired unification between the
inflaton and Dark Matter for free.

\end{abstract}

%\pacs{}

\maketitle

\section{Introduction}

Inflation has now become part of the standard model of
 cosmology  \cite{inflationarycosmology}. With new data coming soon,
and in particular with
the Planck satellite mission, the different inflationary models
will need to pass a strong test.
 Existing  observational experiments involve measurement on CMB \ci{wmap} or large scale structure "LSS" \ci{sdss}
or supernovae SN1a \ci{sn}, and new proposals are carried out.
It has been established that our universe is flat and dominated at present time by Dark Energy "DE" and Dark Matter "DM" with
$\Omega_{DE}\simeq 0.73$,  $\Omega_{DM}\simeq 0.27$ and curvature $\Omega_{k}\simeq-0.012$ \ci{wmap}.
Inflation sets up the initial perturbations from which gravity forms the large scale structures ,
including  Baryon Acoustic Oscillation,  in the universe \ci{sdss}.
Structure formation  requires
the existence of Dark Matter and therefore the nature and dynamics of inflation and  Dark Matter are
essential building blocks to understand the current observations.

Inflation is  associated with a scalar field, the "inflaton", and
the energy scale at which inflation occurs is typically of the
order of $E_{I}= 10^{16} GeV$ \cite{inflationarycosmology}  but it
is possible to have consistent inflationary models with $E_I$ as
low as  $O(100) MeV$\cite{lowinflation}.  On the other hand
Dark Matter is described by an energy density which redshifts
as $\rdm\sim a(t)^{-3}$ and is described by particles where
its mass $m\gg T$, with $T$ its temperature. These particles
can be either fermions or scalar fields. In the case of scalar
fields the scalar potential must be $V(\phi)=m_o^2\phi^2/2$
and independently on the value of the its mass  the classical
equation of motion  ensures that $\rp\sim V \sim a(t)^{-3}$.

Here, since we want to unify inflation with Dark Matter we will
assume that DM is made out of the same field which we denote by  $\phi$.
A scalar field can easily give inflation and DM if
the potential is flat at high  energies and
at low energies the potential approaches the limit $V(\phi)\rightarrow  m_o^2\phi^2/2$ \ci{mio.Q}.
However, most of the time our universe was  dominated by radiation. Therefore, any
realistic model must not only explain the two stages of inflation and Dark Matter
but must reheat the universe  and allow for a long period of radiation domination.
Typically the inflaton decays while it oscillates around
the minimum of its  potential \ci{inflationarycosmology}.
If the inflaton decay is not complete or efficient
then the remaining energy density of the inflaton after reheating
must be fine tuned to give the correct amount of Dark Matter \ci{inf-dm}.

To reheat the universe  we couple $\phi$
to a relativistic field $\vp$ via an interaction term $\lin$. This field $\vp$
may be a standard  model "SM" particle,
as for example neutrinos, but it could also be an extra relativistic particle not contained in the SM.
Here we will assume for simplicity that $\vp$ is an auxiliary  scalar field and
extra relativistic degrees of freedom are fine with the cosmological data \ci{rel.d}.
After inflation the interaction term between  $\p$ and $\vp$ that we use is the standard
 $V_{int} \sim g^2\,\phi^2 \,\vp^2/2$. This process has been widely studied \ci{preheat,preh.tri1,preh.tri2,num.reh} and we
 know that  after a period of preheating and parametric resonance the term $V_{int}$ dominates
 over the inflationary potential $V(\p)$ and one produces $\vp$ particles with $n_\vp\approx n_\p$,
and there is then not a complete decay of $\p$ \ci{preheat} and has also been confirmed an numerically calculated in lattice \ci{num.reh}.
To achieve a efficient decay one needs
to couple $\p$ or $\vp$ to other fields using a trilinear term \ci{preh.tri1,preh.tri2}. We introduce a trilinear
term $h\,\vp \,\overline{\psi }\psi$ between $\vp$ and  fermion fields $\psi$  and we obtain an efficient
decay for $\vp$ and $\p$ which stops at $\Op \approx 10^{-23}$. At a much later stage we show that the
fields $\vp$ and $\p$ are produced using the same interaction terms as for the reheating process.
This is a main difference between other
unification schemes where one needs to fine tune the amount of Dark Matter at the end of reheating and initial
radiation domination epoch. However in our scheme the regeneration of $\p$  takes place
naturally  without fine tuning and,
once the field $\p$ becomes non-relativistic, when its mass  $m_o>T$, $\p$ will decouple since
$n_\p$ is exponentially suppressed. The correct amount of Dark Matter
determines the cross section of Dark Matter at decoupling (as any WIMP particle)
giving a constraint between the coupling $g$ and the mass $m_o$.

\begin{figure}[tbp]
\begin{center}
\includegraphics*[width=8cm]{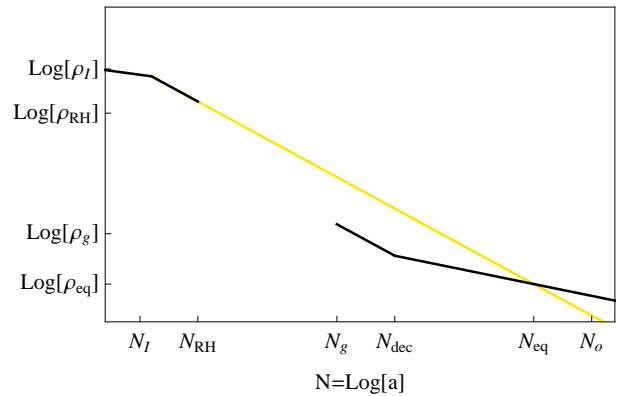}
\caption{We show the evolution of the inflaton-Dark Matter field
$\phi$ (black) and of the relativistic (yellow) energy densities.
Notice that between $N_{RH}$ and $N_{gen}$ there is no
$\rp$ (no $\phi$ particles) and that the evolution of $\rp$ goes from relativistic
to Dark Matter type at $N_{dec}$.}
\label{fig1}
\end{center}
\end{figure}

We will work out the inflation-Dark Matter unification through a simple
example and we show in Fig.\ref{fig1} the evolution of the energy densities  $\rho_\phi$ and radiation.  Of course, the whole scheme is much more general and
other inflaton-Dark Matter potentials  $V(\phi)$ or interaction terms may be used,
however in all cases $V(\phi)\rightarrow m_o^2\phi^2/2$ at late times to have Dark Matter.
The unification scheme  we present  here has four free parameters,
two for the scalar potential $V(\p)$ given by  the inflation parameter $\lm$ of the quartic term
and the mass  $m_o$ given by the quadratic term in $V(\phi)$. The other two parameters
are the  coupling $g$ between the inflaton $\p$ and a scalar filed $\vp$ and the coupling
$h$ between  $\vp$ with standard model particles $\psi$ or $\chi$.
Density perturbations normalized to COBE fixes the value for $\lm$
and the correct amount of Dark Matter
determines the cross section of Dark Matter at decoupling (WIMP particle)
gives a constraint between $g$ and $m_o$, leaving
only the mass of the Dark Matter particle $m_o$ as a free parameter.
However there is a lower limit for $m_o$ from Hot and Warm Dark Matter
with $m_o\geq O(10KeV)$ \ci{warm}   and an upper limit  $m_o\ll E_{h}\simeq 10^6\,GeV$,
so that $\phi$ may be produced. We will show that the
same couplings that give the inflaton decay gives
the Dark Matter regeneration at low scales and sets in combination with
$m_o$ the WIMP decoupling cross section.
These four  parameters are already present in models of inflation and reheating
process \ci{preheat}. This implies that our unification scheme does not increase the number
of parameters and it accomplishes the desired unification between
inflaton and Dark Matter for free.

We organize the present work as follows: In Sec.(\ref{gf}) we present
the general framework, including the scalar potential $V$ and the interaction terms.
In Sec.(\ref{secvint})  we discuss the different interaction terms
for the inflaton decay, the regeneration of the fields $vp$ and$\p$ at a much
later time  and the decoupling of $\p$ as a WIMP particle. Finally we give
the  summary and conclusion in Sec.(\ref{conc}).

\section{General Framework}\la{gf}

Our starting point is a flat FRW universe with the inflaton-dark
matter field $\phi$ coupled to a   scalar $\vp$. We take the
lagrangian $L=L_{SM}+\widetilde{L}$, where $L_{SM}$ is the
standard model SM lagrangian,
\be
\widetilde{L}=\fr{1}{2}\pu
\phi \pU \phi + \fr{1}{2}\pu \vp \pU \vp -V(\phi)-B(\vp)+
\lin(\phi,\vp,SM),
\ee
$V(\phi), B(\vp)$ are the scalar
potentials for $\phi,\vp$ and $\vin=-\lin$  is the interaction
potential. The classical evolution of $\phi$ and $\vp$ are given
by the equations of motion, \bea
\ddot\phi+3H\dot\phi+V'+ \vin' &=& 0\\
\ddot\vp+3H\dot\vp+B_\vp+ \vip &=& 0
\eea
with $H^2\equiv(\dot a/a)^2= \rho/3$, a prime denotes derivative w.r.t. $\phi$,
$B_\vp\equiv \pp B/\pp \vp$ and we take natural units $m_{pl}^2=1/8\pi G\equiv 1$.

There are many potentials $V(\phi)$ that lead to an inflation
epoch at early times and as Dark Matter at late times. Inflation with a single scalar field can be
classified in small or large  field models \ci{sf.lf.inf}.
Small fields are potentials that inflate for values of the inflaton
$\phi\ll m_{pl}$ as in new inflation models, e.g. $V=V_o(\phi^2-\mu^2)^2$,
while large field models are when inflation occurs for $\phi> m_{pl}$ as
in chaotic models, e.g. $V=V_i\phi^n$.
Here we will assume the  simplest
chaotic   potential because we are  interested in showing how
the inflaton-Dark Matter unification scheme takes place in the context of
$\p$ regeneration process. However, more general potentials and interaction
terms may be used in the context of Dark Matter regeneration scheme presented
here.

We take as the inflaton-Dark Matter potential
\be\la{v}
V(\phi)= \fr{1}{2}m_o^2\phi^2 +\fr{1}{4}\lm\p^4
\ee
and we will assume a massless $\vp$, with   $B(\vp)=0$, so that $\vp$ may have only
an effective mass due to its interactions with other fields. The
interaction term that we choose is as standard four leg between $\p$ and $\vp$,
which has been widely studied in the literature in the context of reheating
and preheating \ci{preheat} and we also couple $\vp$ to standard model "SM" particles $\chi,\psi$,
since at high energies the universe must be dominated by SM particles. The interaction we use is
\be\la{vint}
V_{int}=\fr{1}{2}g^2\,\phi^2 \,\vp^2+\fr{1}{2}h^2\,\vp^2\chi^2+ h\,\vp \,\overline{\psi }\psi
\ee
where $g,h$ are constant couplings.  The mass of the different fields are
\bea\la{mp}
\mp^2 &= & V''(\phi) + \vin''=m_o^2+ 3\lm\p^2+ g^2\vp^2,\\
\la{mvp}
m_\vp^2&= & \fr{d^2B}{d\vp^2}  + \fr{d^2\vin}{d\vp^2}=g^2\p^2\\
m_\chi^2 &= & h^2\,\vp^2\\
m_\psi&= & h\,\vp
\eea
and are in general field and time dependent  quantities and we took
$B(\vp)\equiv 0$. We will discuss
the phenomenology  of eqs.(\ref{v}) and (\ref{vint}) in  Sec.\ref{secvint}.

\subsubsection{Inflaton  Potential $V(\p)$}

The potential $V(\phi)$  must satisfy at the inflation scale the slows
roll conditions $|V'/V|<1, |V''/V|<1$ and the constraint on the energy
density perturbation normalized to COBE \cite{inflationarycosmology}
\be\la{dr}
 \fr{\delta \rho}{\rho}= \fr{1}{ \sqrt{75\pi^2}}\fr{V^{3/2}}{V'}=1.9\times 10^{-5}.
\ee
For example  for the chaotic potential   $V_4\equiv \lm \phi^4$ inflation occurs
for $\phi>\phi_e=\sqrt{8}m_{pl}$ and the dimensionless parameter $\lm$
and the scale of inflation $E_I\equiv V(\phi_e)^{1/4}$ are
 \be\la{v4}
V_4\equiv \fr{1}{4} \lm \phi^4,
\ee
\be\la{lm}
 \lm= 2.5\times 10^{-13},\;\;\; E_I=3\times 10^{15} GeV.
\ee
If we take  $V_2\equiv m_o^2\phi^2/2$ as the inflationary potential then
one has $E_I=6\times 10^{15} GeV$ with  a mass $m_o\simeq 10^{13} GeV$. However, since  we are interested
that $\p$   accounts for Dark Matter at low energies we need a much smaller mass  $m_o\ll 10^{13} GeV$  (c.f. eq.(\ref{gg})),
since we have to produce them for $T< E_h\simeq 10^6\, Gev$, and with $m_o=O(GeV)$
the quadratic term is subdominant compared to the quartic term in the potential
in eq.(\ref{v}) at the inflationary epoch.

\subsubsection{Dark Matter  Potential $V(\p)$  }

The field $\p$ should give us Dark Matter at low energies, and its energy density $\rp$
must then  redshifting as $a^{-3}$,  with an equation os state "EOS"
$\wp=0$.  A scalar field  with only gravitational interactions with other fields
and with a potential $V(\phi)=V_i\phi^{n}$ redshifts as $a^{-3(1+w)}$ with $w=(n-2)/(n+2)$ (for n-even) and only for
$n=2$ do we have a energy density redshifting as matter, $V_2\propto \phi^2 \propto a^{-3}$
while for $V_4\propto \phi^4 \propto a^{-4}$ and $w=1/3$ \ci{mio.Q}.
Therefore, the scalar potential $V(\phi)$ must have  the limit at low energies
\be
V(\phi) \rightarrow V_2 \equiv \fr{1}{2}\mmpo \phi^2
\ee
with $\mpo$ a constant mass term as in eq.(\ref{v}). The constraint on $V_2$
is that at present time  (from here on the subscript $o$ gives present time quantities)
\be
\rp(t_o)=\fr{1}{2}\dot\phi_o^2+V_2(t_o)=2V_2(t_o)=m_o^2\phi^2_o=\rdmo
\ee
where we used that the pressure vanishes $p= \dot\phi^2/2- V_2=0$ and
$\rdmo$ is the present time Dark Matter density.  At low energies the mass of the Dark Matter
particle is given by $m_o$ and CDM requires a  mass of $\phi$ to be $m_o\geq O(GeV)$
while warm DM has a smaller mass with $m_o> O(10-100)\, keV$ \ci{warm}.

\subsubsection{Potentials $V_2=V_4 $  }

The potential $ V(\phi)= V_2+ V_4=  \fr{1}{2}m_o^2\phi^2+\fr{1}{4}\lm \phi^4$
is dominated by the quartic term if the oscillations of $\p$ are larger than
\be\la{f24}
\phi\geq  \phi_{24}\equiv \sqrt{\fr{2}{\lm }}\; m_o =2.8\times 10^{6}\, m_o
\ee
and  we define the epoch  where $V_2(\phi)=V_4(\phi)$ at  $\phi= \phi_{24}$ and  $V_{24}=V_2=V_4$   with
\be\la{v24}
E_{24}\equiv V_2(\phi_{24})^{1/4} =  \le(\fr{m_o^2 \phi_{24}^2}{2}\ri)^{1/4}= \fr{m_o}{\lm^{1/4}}=1.4\times  10^{3}\,m_o
\ee
where we have taken $\lm=2.5\times10^{-13}$ from eq.(\ref{lm}) in the r.h.s. of eqs.(\ref{f24}) and (\ref{v24}).
The potential  $V_4$ dominates at high energies, during inflation,
while $V_2$ at low energies, when Dark Matter prevails.

\section{ Interaction Term $V_{int}(\p,\vp,SM)$}\la{secvint}

The aim of this section is to show how a single scalar field $\p$ can account
for inflation at early times and as Dark Matter at a late time, without any fine tuning
of the parameters. In order to achieve this we must have a sufficient decay of the $\p$
field to reheat the universe with SM particles and at a much later stage to regenerate $\rp$
such that it gives the Dark Matter today. We will show that  the interaction term in eq.(\ref{vint})
does indeed allow
for a  sufficient decay of $\p$ at high energies, but at the same time it will
allow for generating $\rp$ at much lower energies without the need to introduce any extra parameters.
The generation of $\rp$ then allows to have $\p$ as Dark Matter.
By sufficient decay we mean that the universe at
the end of reheating and the beginning  of radiation dominated universe has at most
a remanent of $\rp$ which is subdominant compared  to the amount of Dark Matter given by
$\rdm=\rdmo(a_o/a)^{3}$ at that time.

We will now discuss the different interactions and decays but first we present our definition
of sufficient decay in more detail.

\subsection{ Sufficient Decay }

The inflaton decay can have a complete decay into other particles giving
a $\rp \simeq 0$ after reheating. However, we may have a remanent of $\rp$
which may or may not be cosmologically relevant, depending on the
value of $\Op$.

Since  the reheating process gives a radiation dominated
universe the dominate energy density is $\rho\sim  a^{-4}$ while a remanent of $\p$ has  $\rp \sim  a^{-3}$,
taking that  $V_2$ dominates over $V_4$ and that the interaction is now subdominant,
i.e. for $\p\ll \p_{24}$ and the evolution of a  non-interacting fluid with $V_2$ is  $\rp \sim  a^{-3}$  .

The amount of Dark Matter at a
 scale  $E_i=E_o(a_o/a_i)$, with $E\equiv \rho^{1/4}$,  is given by
\bea\la{cdm}
\fr{\Omega_{dm i}}{\Omega_{r i}}
&=&\fr{\rp}{\rho_r}= \fr{\rpo}{\rho_{ro}}\le(\fr{a_d}{a_o}\ri)=
\fr{\Ompo}{\Omega_{ro}} \le(\fr{E_o}{E_d}\ri) \nonumber  \\
&=&  6.3\times 10^{-23} \le(\fr{10^{13} GeV}{E_i}\ri)
\la{opi}\eea
with $\Opo=\Odmo=0.22$, $ \Omega_{ro}h^2= 4.15\times 10^{-5}$ the present time relativistic energy density
and $E_o=2.4\times 10^{-13} GeV$.  If after reheating we have a reheating energy $E_i$ with  $\Op(E_i)\ll \Omega_{dm i}$
then we say the decay was sufficient. A sufficient decay would clearly not allow
$\rp$  to account for Dark Matter since it would not give the correct amount of DM. Of course having
$\Op(E_i)=\Omega_{dm i}$ giving the exact amount of Dark Matter at a high energy $E_i$
would require a fine tuning in the parameters of the model and we want to avoid this scenario.

We can estimate the amount of Dark Matter required at  $E_i=E_{24}=m_o/\lm^{1/4}$ and  using  eq.(\ref{cdm}) we find
\be\la{odm24}
\Omega_{dm }(E_{24})=4.5\times 10^{-13}  \le(\fr{GeV}{m_o} \ri)
\ee
where we have taken eq.(\ref{lm}) and $\Omega_{r i}\simeq 1$.
If the $\p$ field is present
at energies $E_{24}$ its classical evolution  requires that  $\Op$ is  fine tuned to given
the value in eq.(\ref{odm24})  to account for Dark Matter,
unless it is produced again at a later stage.

\subsection{Inflaton Decay and Universe Reheating }\la{reh}

Let us now discuss the interaction term $g^2\p^2\vp^2/2$   between $\p$ and $\vp$
with a scalar potential $V(\p)$ given in eq.(\ref{v}).
This interaction has been widely studied in the context of preheating
and reheating \ci{preheat}. We will present here the main results and we refer to the original work
for details \ci{preheat}.
Soon after inflation, the inflaton field $\p$ oscillates with large amplitude $\p\gg \pd$ and the
$\lm\p^4$ term in the potential $V$ dominates.
The first stage of reheating is a  period of preheating and parametric resonance with
$\phi$ producing $\vp$ particles  in a very efficient process and in a short amount of time \ci{preheat}.
However, the back reaction of $\vp$ is important and  one ends up with an  incomplete decay
of $\p$ \ci{preheat}. In fact it is  the term $g^2\p^2\vp^2/2$  that dominates over $V_4=\lm\p^4/4$,
since the coupling $\lm\simeq 10^{-13} \ll  g^2$,   and one ends with  $n_\p \approx n_\vp$
and $\phi^2\approx \vp^2$. So, taking only the interaction term $g^2\p^2\vp^2/2$ we cannot achieve a complete
decay of $\p$  \ci{preheat}.

In order to have a complete decay one can introduce trilinear terms such  as
$h^2\sigma \p \chi^2$ or $ h \p \overline{\psi }  \psi$, with $\sigma$ an energy scale which can be related to
the v.e.v. of the field $\p$ \ci{preh.tri1,preh.tri2}.  These trilinear terms do produce a complete
decay of $\p$  and we will end up with a universe dominated by SM particles (if $\chi,\psi$ are
SM particles or coupled to the SM). The dominant process at late time is the perturbative decay of $\p \rightarrow \psi\psi$ into
 fermions which dominates over the scattering $\p\p\rightarrow \vp\vp$  \ci{preh.tri1,preh.tri2}.
This decay takes place as long as the decay rate $\Gpff= h^2 m_\p/8\pi$ is larger than $H$ and the mass of $\p$.
satisfies $m_\p > \sqrt{2}\, m_\psi=\sqrt{2} \,h|\p|$ where $m_\p$ is the mass of the scalar field. Even though,
we can reheat the universe with the trilinear term between $\p$ and the fermions $\psi$, this same term will not
 allow $\p$ to play the roll of Dark Matter at late times. This is because the mass of $\p$ at low energies is
 given by the constant $m_o$ and we would then  have a constant decay rate $\Gpff= h^2 m_o/8\pi$, but since $H$ decreases,
 we will have $\Gpff/H \gg 1$ and  the decay  will not stop and $\p$ will decay completely.

In order to avoid a complete decay of $\p$ we choose to couple $\vp$ to fermions $\psi$ with a trilinear term $ h \vp \overline{\psi }  \psi$ as in eq.(\ref{vint}). In this case it will be $\vp$ decaying  into $\overline{\psi}  \psi$ with a decay rate
\be\la{gvff0}
\Gvff= \fr{h^2 m_\vp}{8\pi}
\ee
 and  masses $m_\vp=g|\p|$ and $m_\psi=h|\vp|$.
The trilinear term  $ h \vp \overline{\psi }  \psi$ allows $\vp$ to decay
into fermions $\psi$  in a perturbative process  if the mass condition $m_\vp^2=g^2\p^2> 2 m_\psi^2=2 \, h^2\vp^2$
and $\Gpff/H > 1$  are satisfied. The mass condition gives the constraint $2h^2\leq g^2$ if we take
 $\phi^2\approx \vp^2$. The decay of $\vp$ will then drag the inflaton field $\p$ in the chain reaction
 $\p \leftrightarrow \vp \rightarrow \overline{\psi } + \psi$.  The decay rate is
\be\la{gvff}
\le(\fr{\Gvff }{H }\ri)^2=\fr{3 h^4 m_\vp^2 \mpl^2}{64\pi^2\rho }\equiv \fr{\rho_{R}}{\rho}
\ee
where we have set $3H^2\mpl^2=\rho$ and we define
\be
\rho_{R}\equiv \fr{3 h^4m_\vp^2 \mpl^2}{64\pi^2}
\ee
and  $\Gvff/H\geq $ requires $\rho<\rho_R$ and therefore $\rho_R$ sets the initial stage of the decay. However,
 $\rho_R$ is a function of $m_\vp^2(\p)$. Let us first consider the case when
$V_4$ dominates the potential $V$, i.e. for $\p\gg \p_{24}$. In this case  we have  $m_\vp^4=g^4\phi^4=4   g^4 V_4/\lm$  and
setting $V_4=\rp(1-\wp)/2\sim \rp/3$, $\wp\sim 1/3  $ we have
\be\la{gvff2}
\le(\fr{\Gvff }{H }\ri)^4= \fr{3 h^8 g^4 \mpl^4 }{1024\pi^4\lm }\fr{\rp}{\rho^2}=\Op \fr{\rho_{Ri}}{\rho}
\ee
with $\Op= \rp/\rho$ and
\be\la{ri}
\rho_{Ri}\equiv \fr{3 h^8 g^4 \mpl^4 }{1024\pi^4\lm }   \simeq \le(\fr{2h^2}{g^2}\ri)^4  \le(\fr{m_o}{GeV}\ri)^{6}  \le(2.5\times 10^{14}GeV\ri)^4
\ee
and we have used $2h^2=g^2$   and eq.(\ref{gg}) in last expression of eq.(\ref{ri}).
We have a decay if  $\Gvff/H>1$ which implies a lower limit for $\Op$ given by   $\Op > \rho/\rho_{Ri}$.  Clearly since
$\Op\leq 1$ the upper limit to the initial decay rate is given by $\rho_{Ri}$ with a decay energy $E_{Ri}\equiv \rho_{Ri}^{1/4}\simeq (2h^2/g^2)(m_o/GeV)^{1/6}10^{14}\,GeV$. The  eq.(\ref{gvff2}) gives a balance between $\rp$
and $\rho$ with $\rp \simeq \rho^2/\rho_{Ri} $ and  $\Gvff/H\simeq 1$, which implies that  $\Op$ dilutes as $\rho$  and the decay does not stop.
If $\rp$ dilutes slower than $\rho^2$ than  $\Gvff/H>  1$ giving a decay but  if  $\rp$ dilutes faster  than $\rho^2$
than  $\Gvff/H<  1$ stopping the decay.

So, let us consider now  the  case when $V_2$ dominates $V$. In this case we can  use $m_\vp^2=g^2\phi^2= 2 g^2 V_2/m_o^2$ with  $\rp=2V_2$,  and eq.(\ref{gvff}) becomes
\be\la{gvff3}
\le(\fr{\Gvff }{H }\ri)^2=\fr{3 h^4 g^2 \mpl^2}{64\pi^2 m_o^2}\fr{\rp}{\rho}=\fr{\Op}{\Omega_{\p Rf}}
\ee
where we  defined
\be\la{oend}
\Omega_{\p Rf}\equiv  \fr{64 \pi^2 m_o^2}{3g^2 h^4  \mpl^2}= \le(\fr{g^2}{2h^2}\ri)^2 \le(\fr{GeV}{m_o}\ri) 3\times 10^{-23}
\ee
and we have take a $2h^2=g^2$  and  eq.(\ref{gg}) to evaluate the last expression in eq.(\ref{oend}).
Notice that $\Omega_{\p Rf}$ is independent of $\p$ and $\rho$, so once the ratio $\Op$ becomes smaller than
$\Omega_{\p Rf}\approx  10^{-23}$ the decay stops. We can estimate the energy at the end of the decay given by $\rho_{Rf}=\rp/\Omega_{\p Rf}$ and if we  take the upper limit for $\rp=2V_2=2V_{24}$ we have
\be\la{rf}
\rho_{Rf}= \fr{3 g^2 h^4 m_o^2 m_{pl}^2}{32 \lm \pi^2}\simeq \le(\fr{2h^2}{g^2}\ri)^2  \le(\fr{m_o}{GeV}\ri)^5 (6.8\times 10^{8} GeV )^4
\ee
We conclude that range of energies where the $\vp$ decay takes place is given by the upper limit $E_{Ri}=\rho_{Ri}^{1/4}$ and the lower limit $E_{Rf}=\rho_{Rf}^{1/4}$ given by eqs.(\ref{ri}) and  (\ref{rf}), i.e.
\bea
E_{Ri}&\equiv &  \rho_{Ri}^{1/4}=\le(\fr{2h^2}{g^2}\ri)\le(\fr{m_o}{GeV}\ri)^{1/6}2.5\times 10^{14}\,GeV \\
E_{Rf}&\equiv & \rho_{Rf}^{1/4} = \sqrt{\fr{2h^2}{g^2}}\le(\fr{m_o}{GeV}\ri)^{5/4}6.8\times 10^8 GeV.
\eea
We can  compared  eq.(\ref{oend}) with the amount of Dark Matter in eq.(\ref{cdm}) for different energies.
For example for  $E_i=E_{Ri}\simeq 10^{14} GeV$ we have  $\Omega_{dm }(E_{Ri})=6\times 10^{-24}< \Omega_{\p Rf}\approx  10^{-23} $
and following the discussion below eq.(\ref{ri}),  as long as $V_4$ dominates there is decay, but as soon as $V$ is dominated by
$V_2$ the decay will stop since   $\Op \leq \Omega_{\p Rf}$.
From eq.(\ref{oend}) for   $E_i=E_{Rf}\simeq 10^9 GeV$ we have  $\Omega_{dm }(E_{Rf})=6\times 10^{-19} $ which is much larger than $\Omega_{\p Rf}\approx  10^{-23}$.
We see that
the decay rate $\Gvff$ gives a sufficient decay of $\p$   and if we want to have $\rp$ as Dark Matter
the field $\p$ must be produced at a later stage.

\subsection{Production of $\p$ and $\vp$ at low scales}

If the field $\phi$ decays completely (or sufficiently) after inflation
then there will no residual $\rp$ left to account for Dark Matter, as
seen in then last section in eqs.(\ref{odm24}) and (\ref{oend}).
In order to produce the field $\phi$ at lower energies we use the same
interaction terms  given in eq.(\ref{vint}), i.e. $g^2\p^2\vp^2/2$ and $ h \vp \overline{\psi }  \psi$,
as for the inflaton decay but at a very different  energy scale and where the properties
of the fields are  different.

We have seen in Sec.\ref{reh} that the reheating scale is of the order
of $T_R\simeq 10^{9}-10^{12} GeV$,   the fields $\vp$   does no longer decay into
$\psi$ and that the fields $\vp$ and $\p$ are subdominant, at best.
So we expect that at  low scales $T\ll T_R$, but still above $T \gg TeV$  and $m_o$,
that the  particles $\vp,\p,\psi$ are all relativistic.
This is certainly true for all standard model particles and since the mass
of $\vp$ depends on the value of $\p$,  $m_\vp^2=h^2\p^2\simeq 0$ for $\rp\sim\p^2\simeq 0$.
When the particles   $\psi,\vp$ and $ \p$, are relativistic   the interaction terms
in eq.(\ref{vint})  give  $2 \leftrightarrow 2$ scattering processes.
The interaction rate for the    $2 \leftrightarrow 2$ perturbative  scattering process for relativistic particles is
\be\la{g22}
\G_{22}=\fr{|M_{ab}|^2 n_a }{32\pi^2 E_a^{2}}=\langle \sigma v \rangle n_a.
\ee
where $|M_{ab}|^2 $ is the transition amplitude of the process, $E_a$   the energy
of  the incident particle and   $\langle \sigma v \rangle$ is the cross section times the relative velocity
$v$  of the initial particles  with
\be\la{ss}
\langle \sigma v \rangle=\fr{|M_{ab}|^2 }{32\pi^2 E_a^{2}}.
\ee
Since the particles are relativistic  the
number density is given by $n_a =g_a \zeta[3]\pi^2T^3/30 = c_nE ^3 $ with  $c_n=g_a\zeta(3)/\pi^2\bar{r}^3,$
$E= \bar{r} T$ and $\bar{r}\equiv(\rho/nT)=\pi^4/30\zeta(3)\simeq 2.7$
and eq.(\ref{g22}) becomes
\be\la{g22r}
\G_{22}=c_{22}\; |M_{ab}|^2 E_a
\ee
with $c_{22}=\zeta(3)/(32 \pi^3 \bar{r}^3)$.
In the case of the interaction term  $ h \vp \overline{\psi }  \psi$ the $2 \leftrightarrow 2$
process given by    $\psi +  \psi \leftrightarrow \vp+\vp$  has $|M_{ab}|^2=h^4 $ and
\be\la{gdr}
\G_{\psi\psi\vp\vp}=   c_{22} h^4 E.
\ee
This interaction takes place if  $\G_{\psi\psi\vp\vp} > H$ and
taking  $H=\sqrt{ \rho_r/(3 m_{pl}^2\Omega_r)} \equiv c_H E^2/ m_{pl}$ we have
\be\la{ghdr}
\fr{\G_{\psi\psi\vp\vp}}{H}=\fr{c_d\, h^4  m_{pl}}{E}\equiv \fr{E_h }{E},
\ee
\be\la{eh}
E_{h} \equiv  c_d h^4  m_{pl}= \le(\fr{2h^2}{g^2}\ri)^2  \le(\fr{m_o}{GeV}\ri)^2 2\times 10^{6} GeV
\ee
where  $c_d\equiv c_{22}/c_H $,  $c^2_H\equiv g_r\pi^2/(90\Omega_r\bar{r}^4)$ and
we have take a $2h^2=g^2$  and  eq.(\ref{gg}) to evaluate the last expression in eq.(\ref{eh}).
We see from eq.(\ref{ghdr}) that as long as all
particles  involved, (i.e. $\psi,\vp$ ),  are relativistic and $E<E_h$ we will produce relativistic
$\vp$ particles and $\psi$ and $\vp$ will be in thermal equilibrium for $E\leq E_R$.
For $E> 10^2 GeV$ we have $g_r\simeq 106, \Omega_r\simeq 1$ and
$c_d\simeq 10^{-4}$.

At the same time, once the relativistic field $\vp$ has been produced
the interaction term  $g^2\p^2\vp^2/2$ that couples $\p$ to $\vp$   will produce relativistic $\p$
particles as long as $T\gg m_o$. This interaction term   also gives a
$2\leftrightarrow 2$ perturbative process and from eq.(\ref{g22r}) with  $|M_{ab}|^2=g^4 $ we have
\bea\la{dbd}
\G_{\vp\vp\p\p}=c_{22} g^4 E ,&&\;\;\;\;\;\; H=\sqrt{\fr{\rho_r}{3 m_{pl}^2\Omega_r}} =c_H E^2,\\
\fr{\G_{\vp\vp\p\p}}{H}&&=\fr{c_d\, g^4  m_{pl} }{E}\equiv \fr{E_{g}}{E}
\eea
with $c_d\equiv c_{22}/c_H $ as in eq.(\ref{ghdr}). The process takes place when $\G_{\vp\vp\p\p}/H>1$
or
\be\la{eg}
E\leq E_{g}\equiv c_d g^4  m_{pl}=\le(\fr{m_o}{GeV}\ri)^2 8\times 10^{6} GeV
\ee
and we have take  eq.(\ref{gg}) to evaluate the last expression in eq.(\ref{eg}).
For $E<E_g, E_h$ as long as  all the particles   are relativistic
they are in thermal equilibrium with    and $T_\phi=T_\vp\simeq T_{\psi} $.
From eqs.(\ref{ghdr})  and (\ref{eg}) we have $E_h=(h^4/g^4)E_g \leq  E_g/4$
since $2h^2\leq g^2$. Therefore,
the field $\p$ is produced as soon as $\vp$ are   produced.
As long as $\p$ is relativistic, i.e. $T\gg m_o$, we will $\Omega_\vp=\Omega_\phi$, but once we reach
the region with $T \approx m_o$ the two fields will decouple
since $n_\phi$ will be exponentially suppressed.

\subsection{Non-relativistic $\phi$ Decoupling: $E_{dec}$}\la{secdec}

If two relativistic particles ($\vp,\p$) are in thermal equilibrium and one
(in our case $\phi$) becomes non-relativistic
then the density number $n_\phi$ is exponentially suppressed and
$\vp$ and $\phi$ decouple. This is just the standard WIMP particle
decoupling. The
transition rate  for a $2 \leftrightarrow 2$ process is given by
eq.(\ref{g22})
\be\la{gdec}
\G_{dec}=\langle \sigma v \rangle n_\phi.
\ee
In order to
have the correct amount of Dark Matter a WIMP must decouple with
a $\langle \sigma  \rangle$ such that \ci{wimp}
\be\la{os}
\Omega_{\phi o} h_o^2=\fr{3\times 10^{-27}cm^3 s^{-1} }{\langle \sigma v \rangle}.
\ee
For the $2\leftrightarrow 2$ transition with $\Omega_{\phi o}\simeq 0.22, h_o=0.7$
eq.(\ref{os}) implies a cross section
\be\la{s}
\langle \sigma  \rangle = \fr{g^4}{32\pi m_o^2}=0.1 pb
\ee
with $pb=10^{-36}cm^2=(5\times 10^{-5}/GeV)^2$ and   $v\simeq c$, giving a value for $g^2$
\be\la{gg}
g^2= 1.6\times 10^{-4} \le(\fr{m_o}{GeV}\ri).
\ee
Eq.(\ref{s}) gives a constraint for $g$ in terms of the mass $m_o$. The freeze out
takes place at $x_F=m_o/T_F\simeq 10$ \ci{wimp}  giving a decoupling constant energy
$E_{dec}$  with
\be\la{edec}
E_{dec}= c_{dec} T_{F}= c_{dec} \fr{m_o}{x_F}\simeq 0.12\; m_o,
\ee
and  $c_{dec}=(\pi^2 g_{rel}/30)^{1/4}\simeq 1$ taking $g_{rel}=g_{sm}+g_\vp=  5.5+1=6.5$
at $E<O(MeV)$.
For energies $E<E_{dec}$ the fields $\phi$ and $\vp$ are no longer coupled
and $\phi$ evolves classically as matter with $\rp=\rpo(a_o/a)^3$.
The constraint on warm Dark Matter sets a lower scale $m_o> O(10-100 keV)$ \ci{warm}.

\subsubsection{Dark Matter Decay?: $E_{Ddm}$}

We have seen that at $E_{dec}$ the field $\phi$ ceases to maintain thermal equilibrium
with $\vp$ through the $2\leftrightarrow 2$ process. However,
the field $\phi$ may decay into $\vp$ since $\mp\gg m_\vp$.
Of course, we do not want $\phi$ to decay since it must account
for Dark Matter. In this case the decay $\p\rightarrow \vp\vp$
has a decay rate $\G_{\p\vp\vp}$ for the interaction term $g^2\p^2\vp^2/2$
is
\be\la{ddm}
\G_{\p\vp\vp}=\fr{g^4\phi^2}{8\pi m_o}.
\ee
 Using that $\phi^2=2V_2/m_o^2=\rp/m_o^2$  and $H^2=\rp/3m_{pl}^2$, in matter dominated region,
 we have
 \be
 \le(\fr{\G_{\p\vp\vp}}{H}\ri)^2=\fr{g^8\rp^2}{64\pi^2 m_o^6 H^2}=\fr{3g^8\rp}{64\pi^2 m_o^6}
\ee
and $\G_{\p\vp\vp}/H<1$ if
\be\la{rp}
\rp< \fr{64\pi^2 m_o^6}{3g^4}=\le(\fr{m_o}{GeV}\ri)^{2/3}(3.6\times 10^{13}GeV)^4
\ee
where we have used eq.(\ref{gg}) on the r.h.s. of eq.(\ref{rp}). Eq.(\ref{rp})
clearly shows that $\p$ will not decay into $\vp$ at energies $E\ll 10^{13} GeV$
where Dark Matter is relevant.

\begin{figure}[tbp]
\begin{center}
\includegraphics*[width=8cm]{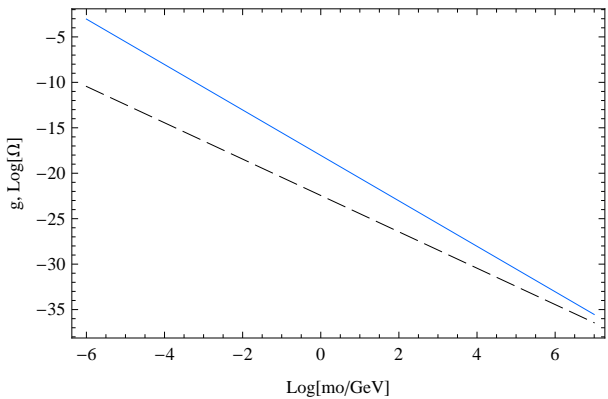}
\caption{We show the dependence of $\Omega_{dm}(E_{Rf})$ and $\Op(E_{Rf})$  (blue, dashed-black, respectively) at $E_{24}$
 as a function of $m_o$. We notice that  $\Omega_{dm}(E_{Rf})\gg \Op(E_{Rf})$ giving  a sufficient inflaton decay.}
\label{fig2}
\end{center}
\end{figure}

\begin{figure}[tbp]
\begin{center}
\includegraphics*[width=8cm]{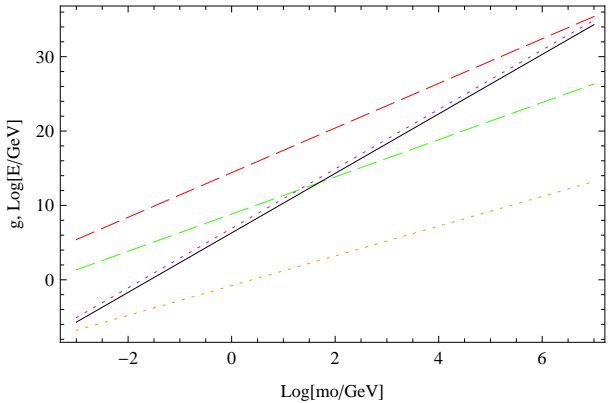}
\caption{We show the dependence of the different energies
densities as a function of $m_o$ in $GeV$ , with
 $E_{Rf}$, $E_{Ri}$, $E_{h}$, $E_{g}$ and   $E_{dec}$  (dashed-red,dashed-green, dotted-blue, black, dotted-orange,
 respectively). Since $E_{h}\sim E_{h}$ both lines overlap and we cannot distinguish  them. }
\label{fig3}
\end{center}
\end{figure}

\begin{table}
\begin{center}
\begin{tabular}{|c|c|c|c|c| }
%\begin{tabular}{|l|l|l|l|l|l|l|l| }
  \hline
  $m_o$  &  $ g^2$ & $\Omega_{dm}(E_{Rf})$ & $\Op(E_{Rf})$ &   $E_{24}$  \\
  \hline
  $10^{-4}$ & $ 1.6\cdot 10^{-8}  $    &   $ 4.6\cdot 10^{-8} $    & $ 3.5\cdot   10^{-19} $    & $ 1.4 \cdot  10^{-1}  $        \\
 $10^{-2}$  &  $ 1.6\cdot 10^{-6}   $    &  $ 4.6\cdot 10^{-10} $    & $3.5\cdot   10^{-21} $   & $  1.4 \cdot 10^{1}  $              \\
  $    1 $ &   $1.6\cdot 10^{-4}    $    &  $  4.6\cdot 10^{-12}  $   & $ 3.5\cdot   10^{-23} $    & $ 1.4 \cdot  10^{3}  $           \\
 $10^{\,2} $  & $ 1.6\cdot 10^{-2}   $     &  $ 4.6\cdot  10^{-14}  $    &$  3.5\cdot   10^{-25}$    & $1.4 \cdot 10^{5}   $       \\
$ 10^{\,4}  $ &  $ 1.6  10^{0}  $                &   $ 4.6 \cdot  10^{-16}  $        &    $ 3.5\cdot   10^{-27} $    &    $  1.4 \cdot 10^{7}  $          \\
  \hline
\end{tabular}
\end{center}
\caption{\small{We show the values of $\Omega_{dm}(E_{24})$,  $\Op(E_{24})$ at the scale  $E_{24}$  in $GeV$ as a function
of $m_o(GeV)$ and the value of the dimensionless coupling $g^2$. }}
\la{table1}\end{table}

\begin{table}
\begin{center}
\begin{tabular}{|c|c|c|c|c| }
%\begin{tabular}{|l|l|l|l|l|l|l|l| }
  \hline
  $m_o$  &  $E_{Ri}$ &  $E_{g}$ &  $E_{h}$ &   $E_{dec}$    \\
  \hline
  $10^{-4}$ & $ 2.5 \cdot  10^{8}  $  & $ 8  \cdot  10^{-2}  $  & $ 2  \cdot  10^{-2}   $ & $ 1.6\cdot 10^{-5}   $      \\
 $10^{-2}$  & $2.5 \cdot     10^{11} $ &    $ 8\cdot  10^{2}  $  & $2 \cdot  10^{2}    $ &  $ 1.6 \cdot 10^{-3} $       \\
  $ 1 $ &  $2.5  \cdot     10^{14} $  & $ 8 \cdot   10^{6}   $   & $2 \cdot   10^{6}   $ & $1.6 \cdot 10^{-1} $        \\
 $10^{\,2} $ & $2.5  \cdot    10^{17}  $  & $ 8  \cdot   10^{10}  $  & $ 2 \cdot  10^{10}   $ & $ 1.6 \cdot 10^{1} $       \\
$ 10^{\,4}  $&  $2.5  \cdot     10^{20} $  & $ 8 \cdot   10^{14}   $  & $ 2 \cdot  10^{14}   $ & $ 1.6 \cdot 10^{3} $        \\
  \hline
\end{tabular}
\end{center}
\caption{\small{We show the values of the energies of  $E_R$, $E_{d}$, $E_{gen}$ and   $E_{dec}$ in $GeV$ as a function
of $m_o(GeV)$.}}
\la{table2}\end{table}

\section{Summary and conclusions}\la{conc}

We have presented a model where inflation and Dark Matter takes place via a single
scalar field $\phi$.
The unification scheme presented here has four free parameters,
two for the scalar potential $V(\p)$ given by  the inflation parameter $\lm$ of the quartic term
and the mass  $m_o$ given in the quadratic term in $V$. The other two parameters
are the  coupling $g$ between the inflaton $\p$ and a scalar filed $\vp$ and the coupling
$h$ between  $\vp$ with standard model particles $\psi$ or $\chi$.
These four parameters are the usual ones used previously for a
scalar potential and interaction couplings describing the inflaton potential and
the reheating process \ci{preheat}. At the same time, the parameters
$m_o$ and $g$ are also present in  Dark Matter models of WIMP particles.
The novel feature here is that without
the introduction of any new parameters  we are able to unify
inflation and Dark Matter using a single field $\p$ that
accounts for inflation at an early epoch  while it gives a
Dark Matter WIMP particle at low energies.

We begin with an inflaton filed $\p$ with a scalar potential $V$ and interaction term $V_{int}$.
After a period of inflation our universe must be reheated and we must account for a long period of radiation
dominated epoch. Typically the inflaton decays while it oscillates around
the minimum of its  potential \ci{inflationarycosmology}.
If the inflaton decay is not complete or sufficient
then the remaining energy density of the inflaton after reheating
must be either subdominant and play no cosmological roll or must be
fine tuned to give the correct amount Dark Matter.
An essential feature is that the transition between the
radiation dominated to the Dark Matter phase is related
to a late production of the scalar field $\phi$, which takes place
naturally  without fine tuning,  instead to a remanent
of Dark Matter energy density   which  needs to be fine tuned.
The scalar potential $V$ has two terms a quartic and a quadratic one. The quartic term
$V_4=\lm\p^4/4$ gives inflation at high energies $E_I\simeq 10^{15}GeV$  and via
the coupling term $g^2\p^2\vp^2/2$ the large oscillations of $\p$ in a preheating stage, with parametric resonances,
produce  large number  of $\vp$ particles but since $\lm\ll g^2$  the term  $g^2\p^2\vp^2/2$
dominates over the potential term   $ \lm\p^4/4$ and one ends up with $n_\vp\approx n_\p$, $\vp^2\approx \p^2$ and
$\Op\approx \Omega_\vp$. To have a complete decay of the inflaton field, we must couple $\p$ or $\vp$
to other fields with a trilinear term. Here we use the coupling $h\vp\overline{\psi}  \psi$ and
the massive $\vp$ decays into two light $\psi$ fields. At the same time, this $\vp$ decay drags
the $\p$ field via the chain reaction
$\p \leftrightarrow \vp \rightarrow \overline{\psi}+\psi$.
We calculate the end of this process and it has $\Op\approx 10^{-23}$ giving a sufficient (efficient) decay. Once
$\p$ and $\vp$ are subdominant the universe has been reheated and we have a standard hot universe filled
with relativistic particles. At much lower energies, the same trilinear interaction
term $h\vp\overline{\psi}  \psi$ produce light $\vp$ particles, via the reaction  $\overline{\psi}+\psi \leftrightarrow \vp +  \vp $
below the energy  $E_h=c_d h^4 m_{pl}$, while
the $g^2\p^3\vp^2/2$ term also produces light $\p$  particles, through the reaction  $\vp +  \vp \leftrightarrow \p +\p $ below  $E_g=c_d g^4 m_{pl}$
and once the $\vp$ have been produced.
We see that   the coupling terms  $g^2\p^2\vp^2/2$ and  $h\vp\overline{\psi}  \psi$ play two different rolls namely
they allow for $\p,\vp$ to decay and reheat the universe at large energy scales, while they also allow to regenerate
these two fields at a much later stages where both fields are relativistic and we have $n_\vp=n_\p$. As a final step, since the mass of $\p$
is different than zero, $m_o\neq 0$, the field $\p$ will become non-relativistic at $T\approx m_o$,
$n_\p$ will be exponentially suppressed and $\p$  decouples from $\vp$, as any WIMP field.
The   amount of Dark Matter today fixes the cross section $\sigma=g^4/32\pi m_o^2$ giving a
constraint between $g$ and $m_o$. We show
in figs.(\ref{fig2}) and (\ref{fig3}) the dependence of the $E's$ on $m_o$
and in tables (\ref{table1}) and (\ref{table2})  we give the values for $10^{-4}GeV<m_o< 10^{4}GeV$.

To conclude,  the unification scheme  we present  here has four free parameters,
two for the scalar potential $V(\p)$ and   two  couplings $g, h$ between the different
fields.   These four  parameters are already present in models of inflation and reheating
process, without considering Dark Matter, and in WIMP models without inflation.
Therefore,  our unification scheme does not increase the number
of parameters and it accomplishes the desired unification between
inflaton and Dark Matter for free.

\begin{acknowledgments}
We would like to thank Luis A. Urena for useful discussions
and comments. We thank for partial support Conacyt Project 80519, IAC-Conacyt
Project.
\end{acknowledgments}

\appendix

\section{Summary of Energies}

We present the different energy scales relevant in the
process of our inflation-Dark Matter unification scheme.
 Concerning the inflaton-Dark Matter unification scheme
we have 4 different parameters $\lm, m_o$ in the potential $V$
and a coupling $g$ between  $\vp$ and $\phi$ and $h$ the coupling
between $\vp$ and $\psi$.

The  energies in eqs.(\ref{ei})-(\ref{ef})  are given in terms of these four parameters.
Inflation fixes one parameter,  $\lm$, and the   amount
of Dark Matter today gives a constraint between
$g$ and $m_o$.  We are left with one single free parameter
which we take it to be the mass $m_o$.  We show
in figs.(\ref{fig2}) and (\ref{fig3}) the dependence of the $E's$ on $m_o$
and in tables (\ref{table1}) and (\ref{table2})  we give the values for $10^{-4}GeV<m_o< 10^{4}GeV$.
We see that $E_{dec}>E_{eq}=O(eV)$ and that the values of $g$
and all other energies are phenomenologically viable. This
implies that it is feasible to implement the inflation-Dark Matter
unification.
We resume the definitions and values of these energies
\bea
g^2&=&(32\pi \langle \sigma \rangle )^{1/2}m_o=1.6\times 10^{-4} \le(\fr{m_o}{GeV}\ri)\\
\la{ei}E_{I} &\equiv & \lm^{1/4} \phi_e=3\times 10^{15}  GeV \\
E_{Ri}&\equiv& \rho_{Ri}^{1/4} = \fr{3 h^8 g^4 \mpl^4 }{1024\pi^4\lm } \\
&=& \le(\fr{2h^2}{g^2}\ri)  \le(\fr{m_o}{GeV}\ri)^{3/2}  \le(2.5\times 10^{14}GeV\ri)\\
\Omega_{\p Rf} &\equiv &   \fr{64 \pi^2 m_o^2}{3g^2 h^4  \mpl^2}\\
&=&  \le(\fr{g^2}{2h^2}\ri)^2 \le(\fr{GeV}{m_o}\ri) 3\times 10^{-23}\\
E_{Rf}&\equiv & \rho_{Rf}^{1/4} \equiv \le(\fr{3 g^2 h^4 m_o^2 m_{pl}^2}{32 \lm \pi^2}\ri)^{1/4}\\
&=& \sqrt{\fr{2h^2}{g^2}}\le(\fr{m_o}{GeV}\ri)^{5/4}\;6.8\times 10^8 GeV.\\
E_{h} &\equiv & c_d h^4  m_{pl} \\
&=&  \le(\fr{2h^2}{g^2}\ri)^2  \le(\fr{m_o}{GeV}\ri)^2 2\times 10^{6} GeV \nonumber    \\
E_{g} &\equiv & c_{d}g^4  m_{pl}\\
&=& \le(\fr{m_o}{GeV}\ri)^2 8\times 10^{6}  GeV \nonumber    \\
E_{dec} &\equiv & c_{dec} T_F = \le(\fr{\pi^2g_{rel}}{30} \ri)^{1/4} \fr{m_o}{x_F}\\
&=& 0.1  \le(\fr{10}{x_F}\ri) \le(\fr{m_o }{GeV}\ri) GeV  \nonumber \\
\la{ef}E_{24}&\equiv &\le(\fr{m_o^2 \phi_{24}^2}{2}\ri)^{\fr{1}{4}} = \le(\fr{4}{\lm}\ri)^{1/4}m_o \\
&=& 1.4 \times 10^3 m_o.\nonumber
\eea

\end{document}